\documentclass[11pt]{cernrep}
\usepackage{graphicx,epsfig,epsf,psfig,amssymb,
rotating,colordvi,helvet,color,subfigure}
\def\gsim{\:\raisebox{-0.5ex}{$\stackrel{\textstyle>}{\sim}$}\:}

\def\B0bar{\overline{B^0}}

\begin{document}
\title{Matrix-element corrections to {\boldmath$gg/q\bar q\to$} 
Higgs in HERWIG}
\author{G. Corcella$^{1,2}$ and S. Moretti$^3$}
\institute{$^1$CERN-TH, Switzerland; $^2$MPI f\"ur Physik, 
M\"unchen, Germany; $^3$Southampton University, UK}
\maketitle
\begin{abstract}
We describe the HERWIG implementation of real matrix-element corrections
to direct Higgs hadroproduction at Tevatron
and Large Hadron Collider (LHC) and compare it to other approaches
existing in literature and
describing the transverse momentum distribution of the Higgs boson.
\end{abstract}

\section{THE HIGGS TRANSVERSE MOMENTUM}

In order to investigate Higgs boson production
via $gg\to$ Higgs (see Ref.~\cite{Kunszt:1996yp}), 
one needs to account for multi-parton radiation
for the sake of performing trustworthy phenomenological analyses
\cite{Cavalli:2002vs,Djouadi:2000gu,Balazs:2000sz}. 
Standard Monte Carlo (MC) algorithms 
\cite{Corcella:2000bw}--\cite{Sjostrand:2003wg} describe parton
radiation in the soft and/or collinear approximation
of the parton shower (PS), but can have regions
of phase space, so-called `dead zones', where no radiation is allowed.
Here, one can however 
rely on higher-order tree-level results, as in this region
the radiation is neither softly nor collinearly enhanced. 
Several methods have been recently suggested in order to match PS
and fixed-order matrix elements (MEs) \cite{Seymour:1994df,Norrbin:2000uu}, 
also including the virtual one-loop terms 
\cite{Frixione:2002ik}--\cite{Dobbs:2001gb}. 

\section{THE HERWIG IMPLEMENTATION}

In this note, we briefly mention that the
same strategy which has already been used to
implement real ME corrections to 
$e^+e^-$ annihilation into quark pairs \cite{Seymour:1992xa}, 
Deep Inelastic Scattering (DIS) \cite{Seymour:1994ti},
top quark decay \cite{Corcella:1998rs} and vector boson hadroproduction 
\cite{Corcella:1999gs} has now also been
adopted for the case of Higgs hadroproduction via gluon-gluon fusion
and quark-antiquark annihilation \cite{cormor},
in the context of the HERWIG event generator 
\cite{Corcella:2000bw,Corcella:2002jc}. That is, the dead zone is here
populated by using the exact next-to-leading order (NLO)
tree-level ME result and the
PS in the already-populated region is corrected using the exact
amplitude any time an emission is capable of being the
{\sl hardest so far}.

\section{NUMERICAL RESULTS AND COMPARISONS}

The MEs squared for the real 
corrections to $gg\to H$ that we have used can be found in
\cite{Baur:1989cm}, where top mass effects are fully included.
The real NLO corrections to $q\bar q\to H$ are instead rather straightforward:
the formulae we used can be read from Eq.~(3.62) of \cite{Moretti:2002eu}
with appropriate Yukawa couplings and crossing.
In the new HERWIG default version, in line with \cite{Corcella:1999gs}, 
ME corrections use
the Higgs transverse mass $m_T^2=q_T^2+m_H^2$ as the scale for 
$\alpha_S$ and for the Parton Distribution Functions (PDFs)
while the $gg,q\bar q\to H$ contributions use $m_H^2$.
We shall also assume that the intrinsic transverse momentum of the
initial-state partons is equal to
$q_{T,\mathrm{int}}=0$, the HERWIG default value.

\begin{figure}
\hspace*{2.5truecm}
{\epsfig{file=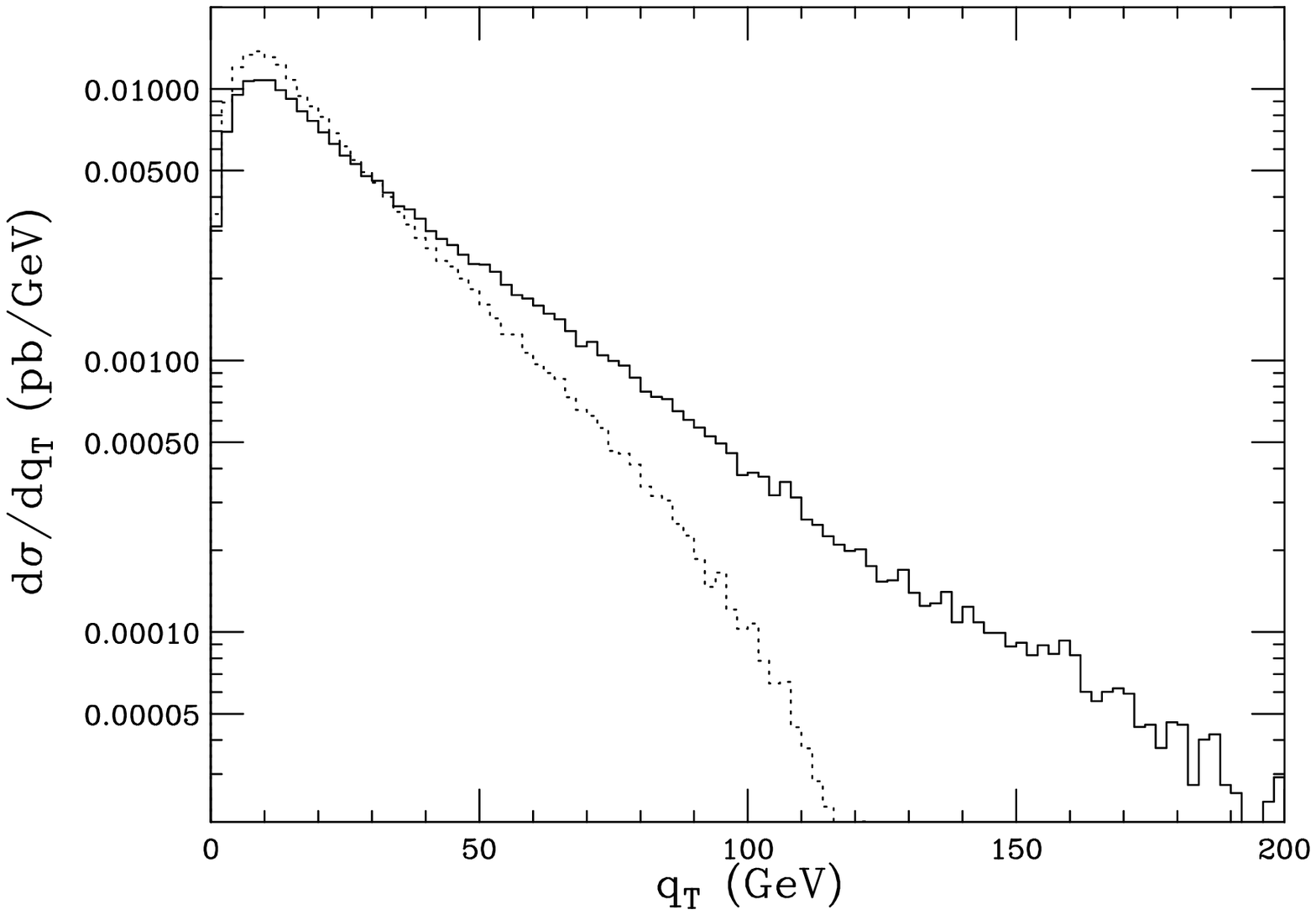,width=5.0cm,angle=0}}
{\epsfig{file=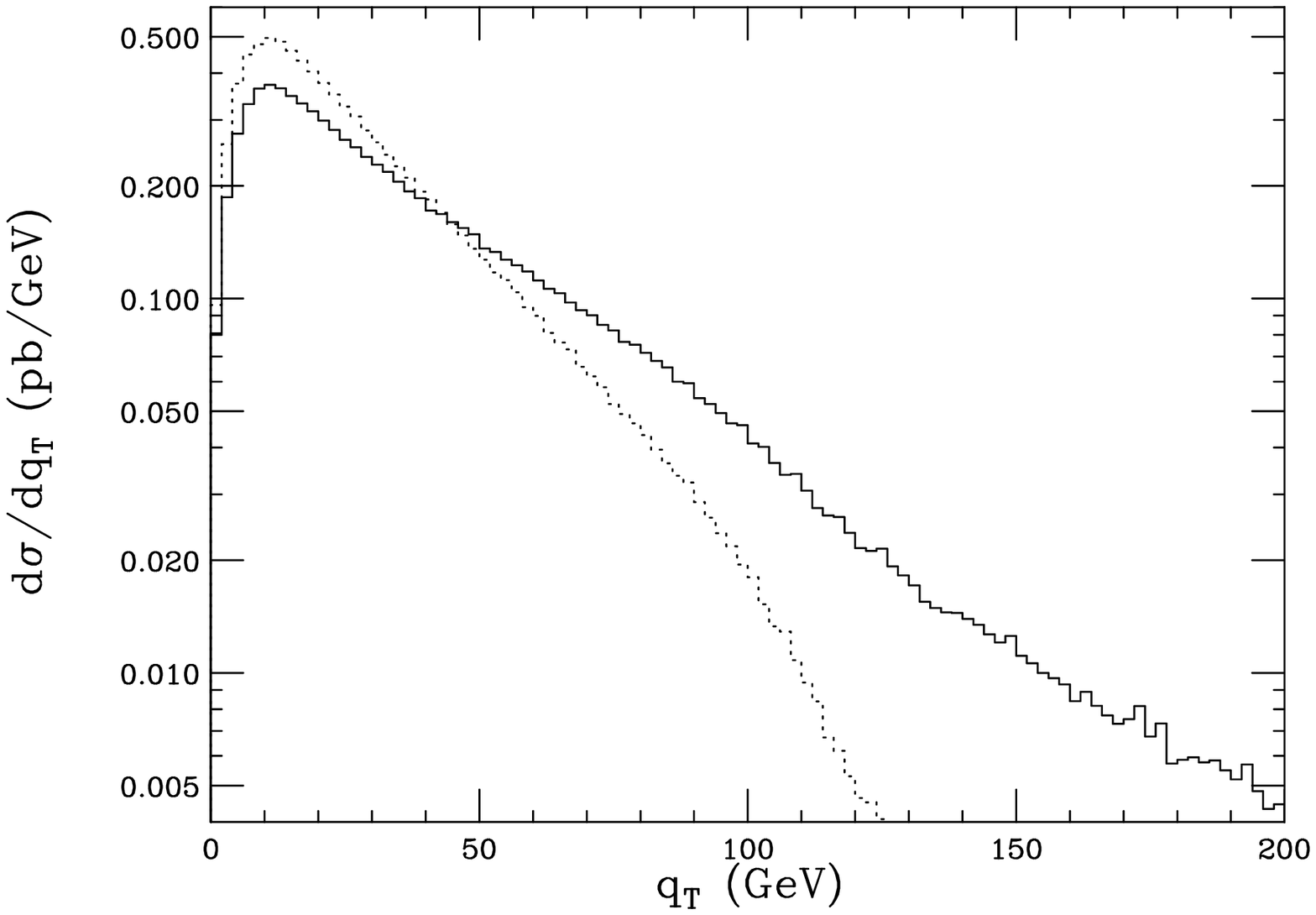,width=5.0cm,angle=0}}
\caption{\small Higgs transverse momentum distribution according to
HERWIG with (solid) and without (dotted) ME corrections,
at Tevatron (left, $\sqrt s_{p\bar p}=2$ TeV) and LHC
 (right, $\sqrt s_{pp}=14$ TeV).
We have set the Higgs mass to $m_H=115$~GeV.}
\label{HW}
\end{figure}

\begin{figure}
{\epsfig{file=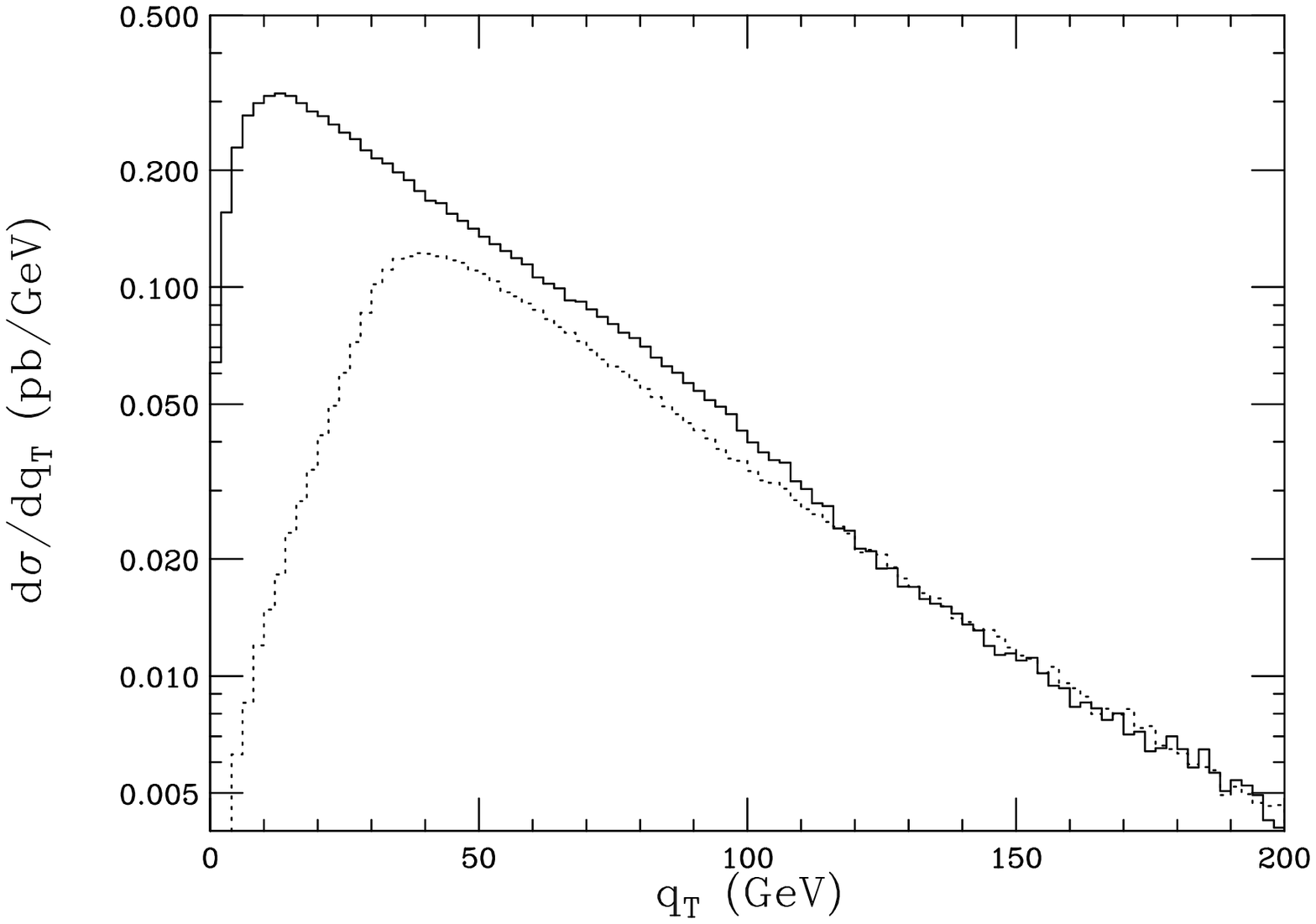,width=5.0cm,angle=0}}
{\epsfig{file=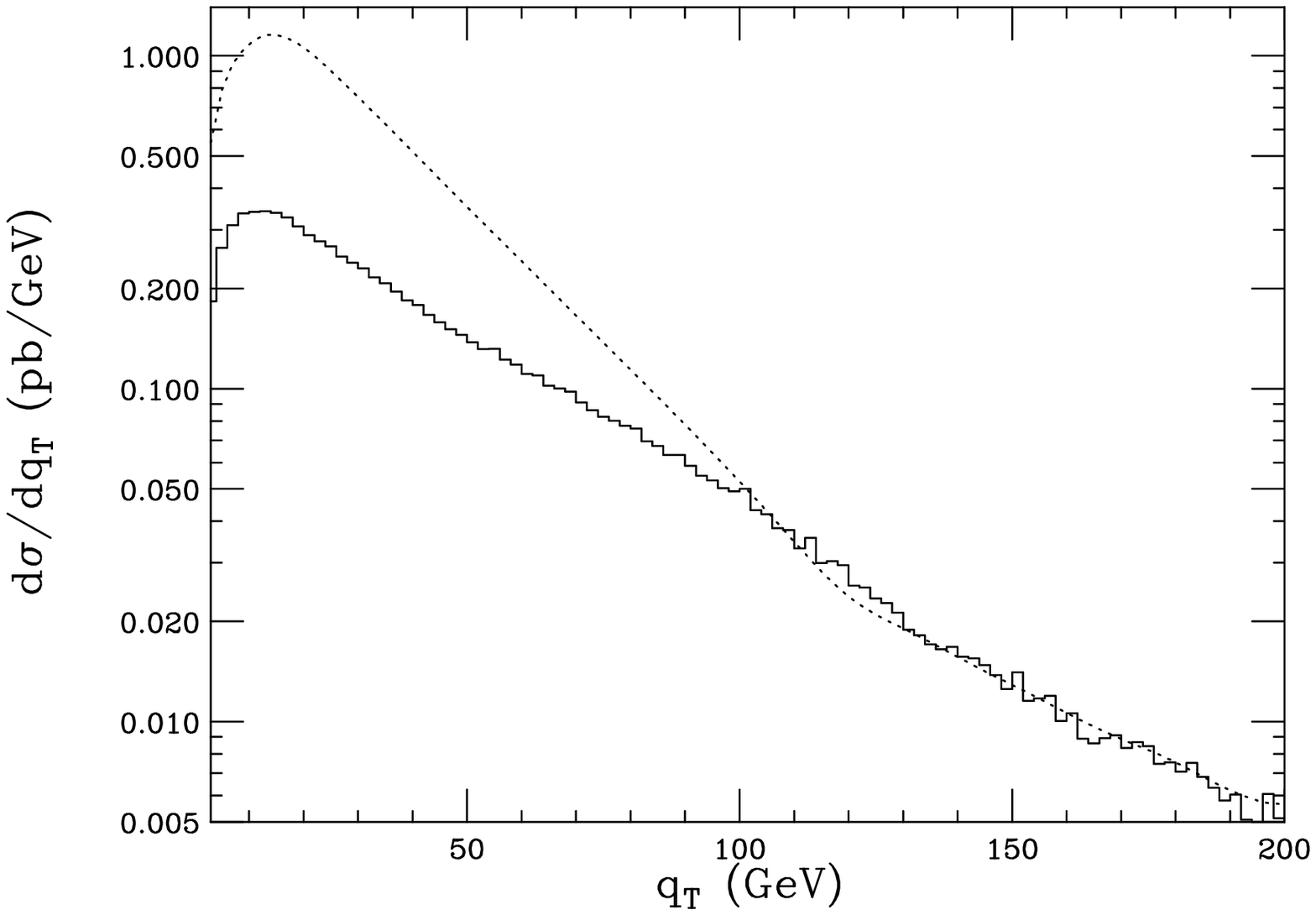,width=5.0cm,angle=0}}
{\epsfig{file=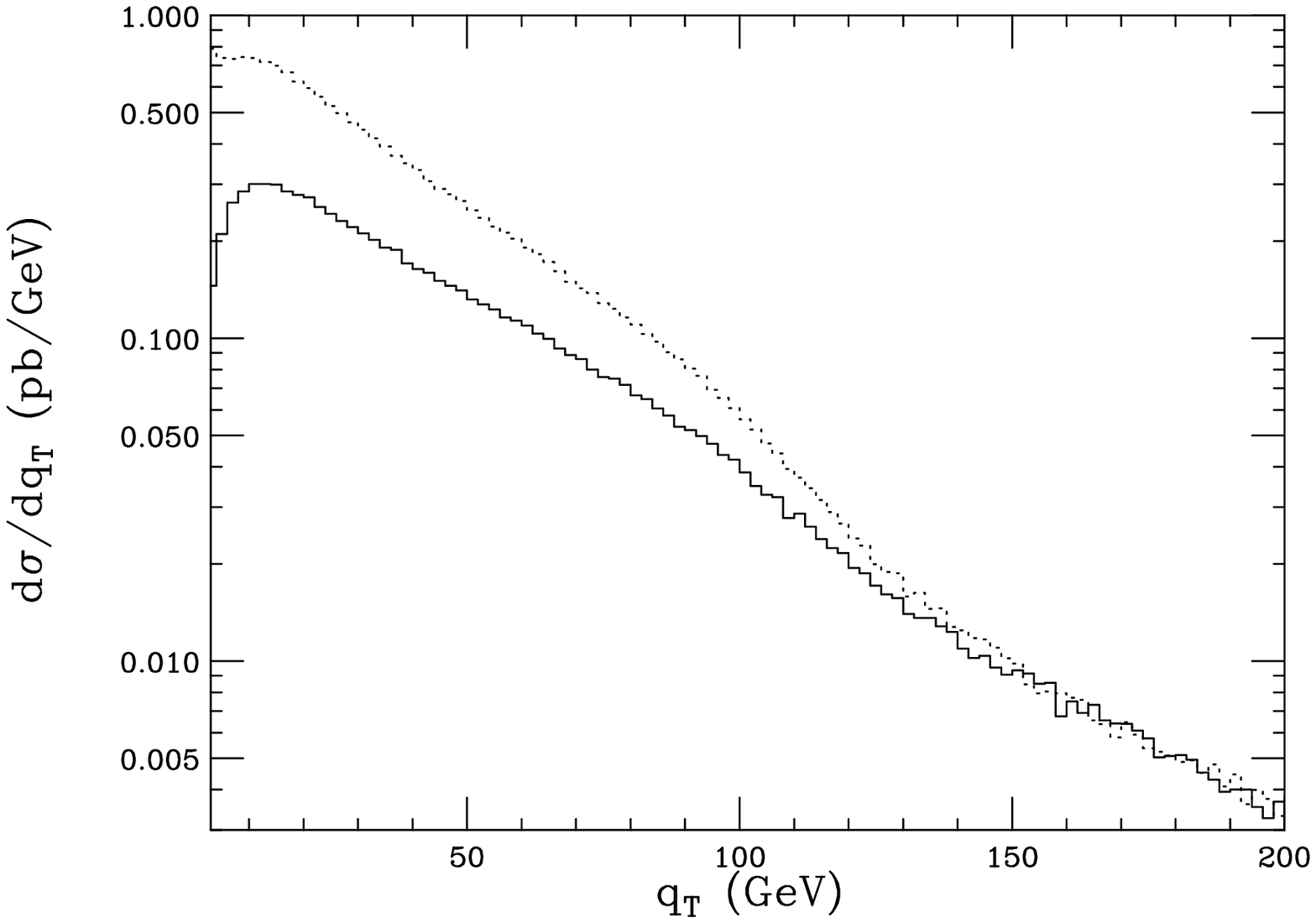,width=5.0cm,angle=0}}
\caption{\small Left: comparison of ME-corrected HERWIG
predictions (solid) to the `$H$ + jets' result from [\ref{bauglo}]
(dotted). Centre: comparison of ME-corrected HERWIG
predictions (solid) to the NLO and resummed calculation of [\ref{bozzi}]
(dotted).
Right: comparison of ME-corrected HERWIG predictions
(solid) to the MC@NLO results from the code described
in Ref.~[\ref{frix}] (dotted). Here,
$q\bar q\to H$ processes have been turned off.}
\label{comparisons}
\end{figure}

By adopting the HERWIG defaults,
we first consider Higgs production at the Tevatron and the LHC within the
MC itself, by plotting the $q_T$ distribution  with (solid histogram) and 
without (dotted) ME corrections: see Fig.~\ref{HW}. 
Beyond $q_T\simeq m_H/2$ the ME-corrected 
version allows for many more events. In fact, one can prove 
that, within the standard algorithm, $q_T$ is constrained to be
$q_T<m_H$. At small $q_T$ the prediction which includes ME
corrections displays a suppression. By default, after the
latter are put in place,
the total normalization still equals
the LO rates. Hence, it is obvious that the enhancement
at large $q_T$ implies a reduction of the number of
events which are generated at small $q_T$ values.

In Fig.~\ref{comparisons} (left plot) 
we present the improved HERWIG spectrum (solid) for the LHC,
along with the result obtained running the so-called
`$H$ + jets' process (dotted), where the hard process is always 
one of the corrections to $gg\to H$. 
In order to perform such a comparison, we have turned 
the $q\bar q\to H$ hard process off, as `$H$ + jets' in HERWIG
does not currently
implement the corrections to quark-antiquark annihilation.
Furthermore, we have chosen $q_{T\mathrm{min}}=30$~GeV for the 
`$H$ + jets' generation.
As expected, at small $q_T$ the two predictions are
fairly different but at large transverse momentum they agree well.

In Fig.~\ref{comparisons} (centre plot) 
we compare the new HERWIG version with the
resummed calculation of Ref.~\cite{Bozzi:2003jy}. For the
sake of comparison with HERWIG, which includes leading logarithms and
only some subleading terms, we use the results
of \cite{Bozzi:2003jy} in the NLL approximation (rather than the default
NNLL one), matched to the NLO prediction. In order for such a comparison to be
trustworthy, we have to make parameter choices similar to \cite{Bozzi:2003jy}:
namely, we adopt a top quark with infinite mass in the
loop and $m_H=125$~GeV, with $\alpha_S$ and PDFs (both from HERWIG
defaults) evaluated at $m_H^2$. While
the normalization (LO in HERWIG, NLO in Ref.~\cite{Bozzi:2003jy}) and
the small-$q_T$ behaviour of the two curves are clearly different, 
the large transverse momentum predictions are in  good agreement, as
in both approaches it is the real NLO ME that
dominates the event generation at large $q_T$.

Finally, in Fig.~\ref{comparisons} (right plot), 
we compare the results of standard HERWIG after ME
corrections with the so-called `MC@NLO' event generator 
(version 2.2) of Ref. \cite{Frixione:2003vm}, the latter 
implementing both real
and virtual corrections to the hard-scattering process, in such
a way that predicted observables (including normalization) 
are correct to NLO accuracy. As version 2.2 of the MC@NLO
includes only the corrections
to Higgs production in the gluon-fusion channel, we  again have
turned the quark-annihilation process off in our routines. 
As observed in the comparison with the resummed calculation, the two spectra 
differ in normalization and at small $q_T$, but agree in the 
large-transverse-momentum region.

\section{CONCLUSIONS}

Between the described implementation and the one available within the 
MC@NLO option,
we believe that HERWIG is presently a reliable event generator for (direct)
Higgs production from parton fusion at hadron colliders both at small and 
large transverse
momentum. In fact, all currently available
ME corrections will play an important role to perform any
analysis on Higgs searches at present and future colliders. In particular,
the option described here may be the most convenient choice for when the
phase space is limited to transverse momentum values such that $q_T\gsim m_H$.

\section*{ACKNOWLEDGEMENTS}

SM would like to thank the 2003 Les Houches workshop organisers
for their kind invitation and the Royal Society (London, UK) for financial
support.

\end{document}